\documentclass[preprint2]{aastex}

\shorttitle{Variables stars in UMa~I}
\shortauthors{Garofalo et al.}

\begin{document}

\title{VARIABLE STARS IN THE ULTRA-FAINT \\
 DWARF SPHEROIDAL GALAXY URSA MAJOR I\altaffilmark{*}}

\author{ALESSIA GAROFALO\altaffilmark{1,2}, FELICE CUSANO\altaffilmark{2}, GISELLA CLEMENTINI\altaffilmark{2}, VINCENZO RIPEPI\altaffilmark{3}, MASSIMO DALL'ORA\altaffilmark{3}, MARIA IDA
MORETTI\altaffilmark{1,2,3}, GIUSEPPINA COPPOLA\altaffilmark{3}, ILARIA MUSELLA\altaffilmark{3}, MARCELLA MARCONI\altaffilmark{3}}

\affil{$^1$Dipartimento di Astronomia, Universit\`a di Bologna,  Via Ranzani 1, I - 40127 Bologna, Italy}
\email{alessia.garofalo@studio.unibo.it}   

\affil{$^2$INAF- Osservatorio Astronomico di Bologna, Via Ranzani 1, I - 40127 Bologna, Italy}
\email{fcusano@na.astro.it, gisella.clementini@oabo.inaf.it}

\affil{$^3$INAF- Osservatorio Astronomico di Capodimonte, Salita Moiariello 16, 
I - 80131 Napoli, Italy}
\email{ripepi@na.astro.it, dallora@na.astro.it, imoretti@na.astro.it, coppola@na.astro.it, ilaria@na.astro.it, marcella@na.astro.it}

\altaffiltext{*}{Based on data collected  at the 2.0 m telescope of the Thueringer Landessternwarte, Tautenburg (TLS), Germany;  the 1.52 m telescope of the INAF-Osservatorio
Astronomico di Bologna, Loiano, Italy;  the 1.54 m Toppo telescope (TT1) of the INAF-Osservatorio
Astronomico di Castelgrande, Italy; and archive data collected with the  Suprime-Cam  of the Subaru Telescope.}

\begin{abstract}
We have performed the first study of the variable star population
of Ursa Major I (\object{UMa~I}), an ultra-faint dwarf satellite recently discovered around 
the Milky Way by the Sloan Digital Sky Survey. 
Combining time series observations in the $B$ and $V$ bands from four different telescopes, 
we have identified seven RR Lyrae stars in UMa~I, of which five are fundamental-mode (RRab) 
and two are first-overtone pulsators (RRc). Our $V$, $B-V$ color-magnitude diagram of UMa~I reaches
V$\sim$23 mag (at a  signal-to-noise ratio of $\sim$ 6) and shows features typical of a single old stellar population.
The mean pulsation period of the RRab stars 
$\langle P_{\rm ab} \rangle$ =0.628, $\sigma = 0.071$ days (or $\langle P_{\rm ab} \rangle$ =0.599, $\sigma= 0.032$ days, if V4, the longest period and brightest variable, is discarded) 
and the position on the period-amplitude diagram suggest an Oosterhoff-intermediate classification for the galaxy. 
The RR Lyrae stars trace the galaxy horizontal branch at an average apparent magnitude  of 
$\langle V(RR)\rangle$ = $20.43\pm0.02$ mag (average on 6 stars and discarding V4), giving in turn a
distance modulus for UMa~I  of (m-M)$_0$ = $19.94\pm0.13$ mag, distance d= $97.3^{+6.0}_{-5.7}$ kpc, 
 in the scale where the distance modulus of the Large Magellanic Cloud is $18.5\pm0.1$ mag.
Isodensity contours of UMa~I red giants and horizontal branch stars (including the RR Lyrae stars identified in this study) show that the galaxy has an S-shaped
structure, which is likely caused by the  tidal interaction with the Milky Way. 
Photometric metallicities were derived for six of the UMa~I RR Lyrae stars from the parameters of the Fourier 
decomposition of the $V$-band light curves, leading to an average metal abundance of [Fe/H]=$-$2.29 dex ($\sigma$=0.06 dex, average on 6 stars)  on the Carretta et al. metallicity scale.
\end{abstract}


\section{INTRODUCTION}

The hierarchical theory of galaxy formation predicts that smaller galactic
  structures merge and dissolve to form the galaxies that we see today.
Since 2004, seventeen  new dwarf companions have been discovered around the  Milky Way (MW)  from the analysis of the Sloan Digital Sky Survey (SDSS) data \citep[see, e.g.,][2010 and references therein]{bel07}. A similar number of new satellites have been discovered also around M31 \citep[see, e.g.,][and references therein]{richardson11}.
The new systems generally have  luminosities fainter \citep[L$_{\rm v} \sim  10^{3} - 10^{4} L_{\odot}$,][]{bel07} than the classical dwarf spheroidal (dSph) galaxies and have thus be named 
ultra-faint dwarfs (UFDs).
They are usually dark matter
   dominated  and spectroscopic analyses have revealed that they contain extremely metal-poor stars with metallicities as low as ${\rm[Fe/H]} = -4.0$ dex \citep{tol09}. In a number of cases they also show an irregular shape probably due to the tidal interactions with the MW.  
 The UFD galaxies being discovered in large numbers around 
the Milky Way (MW) and M31 galaxies are perhaps the  best candidates for the  lone survivors of the cannibalistic construction of the two giants spirals of the Local Group (however, see 
\citealt{pawlowki02a,pawlowki02b}, for challenges to the cosmological accretion scenario of galaxy formation).

The majority of the UFDs studied so far for variability have been found to contain pulsating variable stars of RR Lyrae type with pulsation characteristics 
 \citep[see, e.g.,][and references therein]{cle12,mus12} 
conforming to the properties of the variables in the MW globular clusters (GCs).
In the MW the GCs that contain RR Lyrae stars divide into two different groups 
according to the mean period  of the fundamental-mode (RRab) RR Lyrae stars  \citep{oos39}, and the number ratios of fundamental to first-overtone (RRc) pulsators ($f_c = N_c/N_{ab+c}$, where 
$N_c$ and $N_{ab}$ are the numbers of first-overtone and fundamental-mode RR Lyrae stars, respectively): Oosterhoff type I (OoI) clusters have $\langle P_{\rm ab}\rangle$ $\simeq$ 0.55 days and $f_c \sim$ 0.17, while 
Oosterhoff type II (OoII) 
GCs have $\langle P_{\rm ab}\rangle$ $\simeq$ 0.65 days and $f_c \sim$ 0.44 \citep{cle01}.
Differences in the mean periods of the RRc stars are also found between the two groups, with the 
OoI GCs having RRc mean periods of  $\sim$ 0.32 days and the OoII GCs having mean periods of $\sim$ 0.37 days  (see, e.g., \citealt{cat09}).
The Oosterhoff dichotomy is related to the metal content as OoII GCs are generally more metal-poor than the OoI GCs, but also 
 the horizontal branch (HB) morphology plays an important role \citep{lee90,lee99,cap00}. The Oosterhoff dichotomy is also present among the MW field RR Lyrae stars as first shown by \citet{bon97} and later confirmed by other studies  \citep[see, e.g.,][]{miceli08,cat09}.  
The study of the Oosterhoff properties of the dwarf satellite companions of the MW plays an important role 
in identifying which satellites may have been the building blocks of the Galactic halo.
Indeed, the Oosterhoff  dichotomy and the existence of an Oosterhoff gap has been observed so far only in the MW, as field and 
 cluster RR Lyrae stars in the $``$bright$"$ classical dSphs surrounding the MW have 
0.58$\leq \langle {\rm P}_{\rm ab}\rangle \leq$0.62 days and fall preferentially into the so-called Oosterhoff gap \citep{cat09,cle10} avoided by the MW 
GCs\footnote{ It remains still unclear whether other large spiral galaxies  do exhibit  an Oosterhoff dichotomy.  
It may be possible that also Andromeda has an Oosterhoff gap (e.g., \citealt{contreras13} and references therein)
but further investigations are necessary.}
Thus, the MW halo cannot have been assembled by accretion of dwarf galaxies resembling 
the present-day bright MW dSph satellites. On the contrary, 
the vast majority of the UFDs are found to contain RR Lyrae stars
 with OoII characteristics 
\citep[see][ and references therein]{cle12,mus12}, thus supporting the hypothesis 
that systems resembling the present-day UFDs might indeed have contributed to the building of the MW halo.

UMa~I  \citep[R.A.$=10^{h}34^{m}44^{s}$,
DEC. $=51^{\circ}55^{\shortmid}33.9^{\shortparallel}$, J2000.0;
 l=$160^{\circ}$, b=$54^{\circ}$;][]{wil05a} is,  along with Willman 1 \citep{wil05b},  the first of the MW UFD satellites to  be discovered by the SDSS.
The galaxy has an half-light radius of 
  r$_{\rm{h}}=11.3 \pm 0.5\arcmin $ \citep{mar08}, that at the distance 
  of 96.8$\pm$ 4 kpc \citep{oka08} corresponds to r$_{\rm{h}}= 318^{+50}_{-39} $ pc. 
According to the absolute magnitude M$_{\rm V} \sim -6.75$ mag estimated by \citet{wil05a}, UMa~I 
    is about 8 times less luminous than the faintest of the $``$bright$"$ classical dSphs, namely, Draco, Sextans and Ursa Minor. 
Based on a $V, V-I$ color-magnitude diagram (CMD) 
reaching $V \sim 25$ mag, \citet{oka08} conclude that UMa~I  contains an old stellar population comparable in age to the Galactic GCs M92 and M15. 
This was recently confirmed by \citet{bro12} whose much deeper CMD ($m_{F814W}$ $\sim 28.5$ mag) 
obtained with the ACS on board the \textit{Hubble Space Telescope (HST)} shows the presence 
in UMa~I of a single stellar population with an age of $\sim$ 13.6 Gyr.
Measuring the velocity dispersion of the brightest stars in UMa~I,
      \citet{kle05} and \citet{mar08} found that the galaxy
        is one of the most dark matter dominated objects in the Universe, 
	with a mass-to-light ratio M/L $\sim 1000$ M$_\odot$/L$_\odot$. \citet{sim07} confirmed the systemic velocity derived in these 
	previous studies using a larger sample of stars,  but derived a lower 
	 dispersion. Nevertheless,  assuming the lower luminosity of UMa~I  
(M$_{\rm V}\sim -5.5$ mag) measured by \citet{bel06}, \citet{sim07} estimated an M/L similar to that 
	 of \citet{kle05} and \citet{mar08}.  In the same work
\citet{sim07} also derived a metal abundance of [Fe/H]=$-2.06\pm 0.10$ dex from spectra of the galaxy red giants, 
later revised  to [Fe/H]=$-2.29\pm 0.04$ dex by \citet{kir08}, and to [Fe/H]=$-2.18\pm0.04$ dex by \citet{kir11}. 
UMa~I  appears to be strongly elongated \citep{mar08}, showing the  signs of a possible tidal interaction with the MW \citep{oka08}.

In this paper we present results from the study of the variables stars populating UMa~I.
%
The paper is organized as follow: observations, data reduction and calibration of the UMa~I photometry are presented in Section 2. 
Results on the identification and characterization of the variable stars, 
the catalog of light curves, and the Oosterhoff classification are discussed in  Sections 3. 
The distance to UMa~I and the metallicity of the RR Lyrae stars, derived from the Fourier analysis of the light curves, are presented in Section 4.
The galaxy CMD is discussed in Section 5 along with the spatial distribution of UMa~I stellar components.  
Finally, a summary of the main results is presented in Section 6.

\section{OBSERVATIONS AND DATA REDUCTIONS}

Time series $V,$ $B$ photometry of UMa~I was obtained in the period 2009-2010 at three different telescopes. 
The first set of $V$ images 
was acquired using the 
Bologna Faint Object Camera (BFOSC) mounted on the 1.52 m 
Cassini Telescope of the INAF-Bologna Observatory 
in Loiano\footnote {See http://www.bo.astro.it/loiano/index.html}.
BFOSC is a multipurpose instrument for imaging and spectroscopy, 
equipped with a 1340$\times$1300 pixel EEV CCD 
with 0.58 arcsec $pixel^{-1}$ scale. The total field of view (FOV)
 is of 12.6 $\arcmin$$\times$13 $\arcmin$ and to fully cover the galaxy 
 we needed two adjacent pointings.
Observations of UMa~I  were performed in two nights in 
2009 February, and in three nights in 2009 March, 
obtaining a total of 21 $V$ images each with 1500 s exposure.
 The second set of 41 $V$ images as well  38 B exposures were obtained
 using the 2.0 m telescope at the 
Thueringer Landessternwarte Tautenburg (TLS)\footnote {See http://www.tls-tautenburg.de/TLS/index.html} equipped
 with a 2k$\times$2k SITe CCD at the Schmidt focus. The total FOV of the camera is
 42$\arcmin$$\times$42$\arcmin$, with a pixel 
 scale of 1.2 arcsec/pixel. Observations were performed 
in 2009 April and May.
A third dataset  consisting of a few $V$ images  was obtained  at  the 1.54 m Toppo telescope TT1 
of the Osservatorio Astronomico di Castelgrande\footnote {See http://www.oacn.inaf.it/tt1.html} using the 
2k$\times$2k SITe Toppo Telescope Scientific Camera (TTSC) which has a useful area of 
2048$\times$2048 pixels and a total FOV of $\sim 13\arcmin$$\times$$13\arcmin$.
These data were collected during one night in 2009 November 
and two nights in 2010 March.
Finally, to increase the data sample and to enlarge  the temporal baseline we complemented our photometry  with $V$-band archive images of UMa~I obtained with 
the  Suprime-Cam at the Subaru Telescope (PI N.Arimoto ID proposals: o05167/o05223). These data were acquired  during 
the nights from  2005 December 31 to 2006 January 3.
The  Suprime-Cam consists of a 5$\times$2 arrays of 2048$\times$4096 CCD detectors
and provides a total FOV of 34$\arcmin$$\times$27$\arcmin$.
 The log of all the observations used in the present study is provided in Table~\ref{t:0}.\\
Images were pre-reduced (bias-subtracted, trimmed, and flat-fielded) using standard routines within IRAF\footnote{IRAF is distributed by the National Optical Astronomical
Observatory, which is operated by the  Association of Universities for Research
in Astronomy, Inc., under cooperative agreement with the National Science
Foundation}. The PSF photometry was then performed using the \texttt{DAOPHOT-ALLSTAR-ALLFRAME} packages  \citep{ste87,ste94}. 
  The  alignment of the images was performed using  \texttt{DAOMATCH}, one 
  of the routines in the  \texttt{DAOPHOT} package, whereas  \texttt{DAOMASTER}
  \citep{ste92} was used to match the point sources. The $B$ and $V$ images were aligned
  using as a reference an image obtained at the TLS telescope, which has the 
  largest FOV. This image was astrometrized using the WCStools available at
\textit{http://tdc-www.harvard.edu/ wcstools/}, allowing us to astrometrize our
catalogs. 
To perform the photometric calibration, as a first step, 
  we  cross-matched our photometric catalog with the SDSS catalog  
  \citep{aba09}. From the SDSS catalog only objects flagged as stars  
   and with good quality of the observations were selected.
A total of 1118 stars were found to be in common between the two catalogs.
  The $g$ and $r$ magnitudes of the SDSS stars were first converted to  standard Johnson  $B$ and $V$ magnitudes 
  using the calibration equations by Lupton (2005) available at \textit{http://www.sdss.org/dr4/algorithms/ sdssUBVRITransform.html}. 
As a last step  the parameters of the photometric calibration were derived 
 fitting the data to the equations $B_{\rm s}-b$= c$_{\rm B}$+m$_{\rm B}\times (b-v)$ and  $V_{\rm s}-v$=c$_{\rm V}$+m$_{\rm V}\times (b-v)$ 
where $B_{\rm s}$ and $V_{\rm s}$ are the magnitudes of  the SDSS stars in the Johnson system,  and $b$ and $v$ are the  magnitudes in our  catalog. 
The fit was performed using a 3 $\sigma$ clippling rejection algorithm. A total of 581 stars were used in the final calibration, 
 with magnitudes ranging from 15 to 23 mag in  $B_{\rm s}$ 
and  from $-$0.1 to 1.6  mag in $B_{\rm s}-V_{\rm s}$.
The final r.m.s. of the fit is of 0.05 mag both in $B$ and $V$. This accuracy is  adequate 
 for our main purpose of identifying the galaxy variable stars.

\begin{table*}
\begin{center}
\caption[]{Log of the observations}
\label{t:0}
\begin{tabular}{l l c c c c c}
\hline
\hline
\noalign{\smallskip}
 Telescope&{\rm ~~~~~Dates}& {\rm Filter}&        N   & Exposure lenght & {\rm FWHM}\\
 	   &		   &             &	      &  (s)           &  {\rm (arcsec)}\\
 	    \noalign{\smallskip}
	    \hline
	    \noalign{\smallskip}
 Cassini&  2009, February 25-26 &    $V$   &        10 & 1500  &2.5-3  \\
        &   2009, March 23-25    &   $V$   &       11 & 1500   &2.5-3  \\ 
	&                          &         &          &        &    &     \\
     TLS &{\rm 2009, April 19-25}     & $B$        & 21 &  900  &2-2.2  \\
         &{\rm 2009 May 18}          & $B$        & 17 &  900  &2-2.2 \\   
         &{\rm 2009, April 19-25}     & $V$        & 24 & 1200 &2-2.2  \\
         &{\rm 2009, May 18}          & $V$        & 17 & 1200 &2-2.2 \\ 
         &                            &            &    &    &  &     \\
  TT1 &{\rm 2009, November 26}     & $V$           & 1  & 1500&2.5  \\  
      &{\rm 2010, March  20}     & $V$               & 3  & 1500&2.5  \\       
      &{\rm 2010, March 23}     & $V$                &  1 & 1500&2.5  \\ 
       &                         &                    &    &  &     \\ 	 
 Subaru&  2005, Dicember 31& $ V$                 & 5  & 10&0.6 \\
            &  2006, January 1-3& $ V  $           & 24 & 10&0.7\\         
           &                        &   	      &    &	&	   \\

\hline
	 \end{tabular}
	 \end{center}
	 \normalsize
	  \end{table*}

\section{IDENTIFICATION OF THE VARIABLE STARS}\label{sec:var}

\begin{figure*}[!t]
\centering
\includegraphics[scale=0.95]{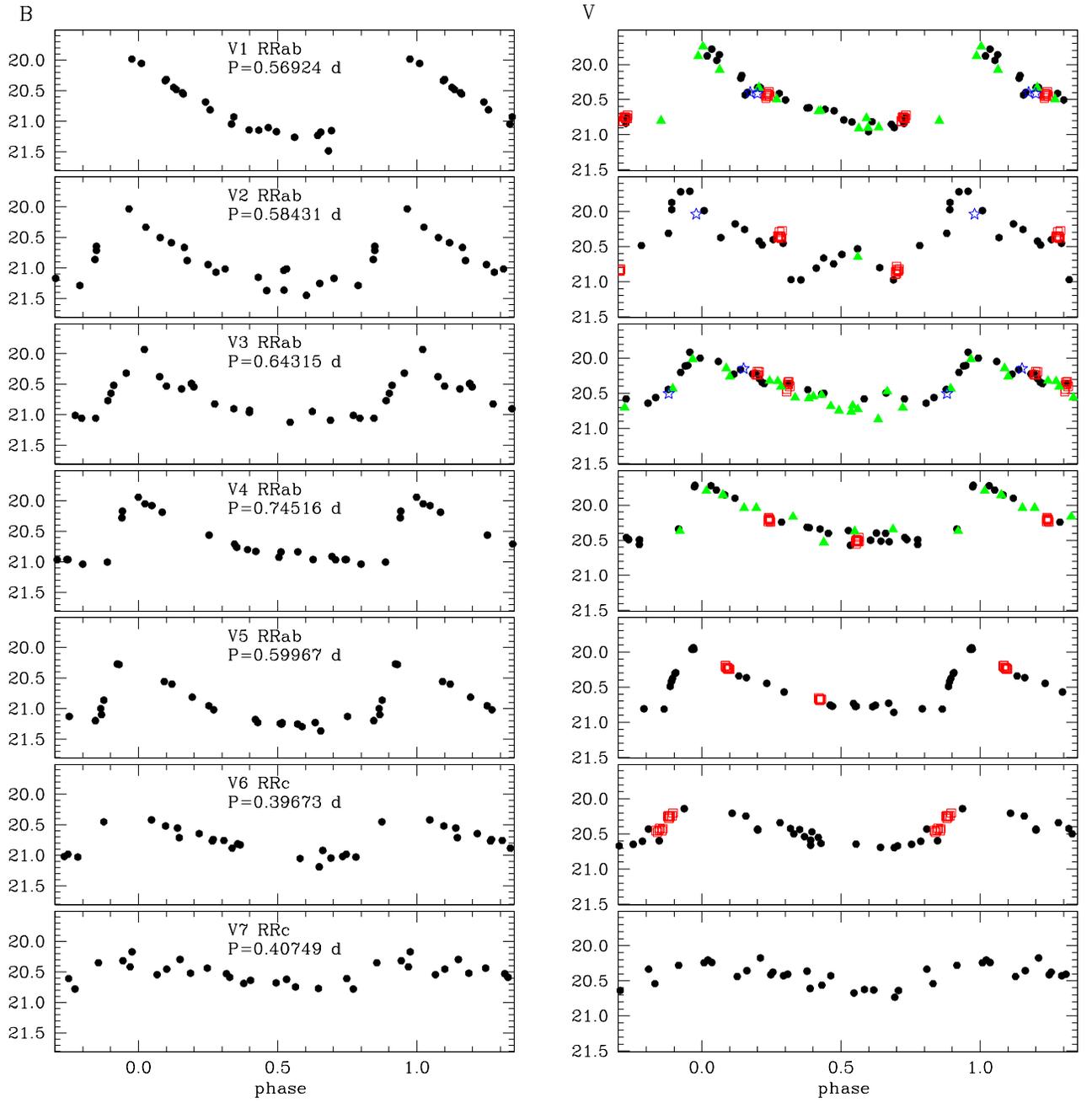}
\caption[]{$B$  \textit{(left panels)} and $V$  \textit{(right panels)}  light curves of the
 RR Lyrae stars we have identified in the UMa~I galaxy. Black dots are TLS data, 
open (red) squares are SUBARU data, (green) triangles are Loiano data, and  open (blue) stars  are TT1 data.} 
\label{fig:curve}
\end{figure*}

Variable stars were identified  on the $V,B$ data separately, using the variability index computed in \texttt{DAOMASTER} \citep{ste94}.
The signal-to-noise (S/N) and the time coverage of the UMa~I data is such that it allows the detection of variable stars  as faint 
as  $V \sim 22$ mag and $B\sim 22.5$ mag, and with periodicities in the range of a few hours to a few days for the $B$ time series, and from 
a few hours to several days for the $V$ data.
A sample of 130 candidate variable stars were identified.
Only candidate variables consistently varying in both passbands were retained. 
 The $V$, $B$ light curves of the candidate variables were analyzed
 with the Graphical Analyzer of Time Series (GRATIS), a private software developed at the Bologna Observatory 
by P.Montegriffo \citep[see, e.g.,][]{clm00} that first performs the period search using the Lomb periodogram \citep{lom76,sca82} and then  the 
best fit of the data with a truncated Fourier series 
\citep{bar63}. The final period adopted to fold the light curves was the one  minimizing 
 the r.m.s. scatter of the truncated Fourier series best fitting the data.
 We confirmed the variability and obtained reliable periods
  for seven variables in UMa~I, all of RR Lyrae type.  
  Five of them are fundamental-mode pulsators, 
  while the remaining two are first-overtone variables. The identification and the properties of the RR Lyrae stars detected in UMa~I are 
  summarized in Table ~\ref{t:1}. 
We have assigned to the variables an increasing number starting from the galaxy center for which we adopted the coordinates by \citet{mar08}. 
Column 1 gives the star identifier, Columns 2 and 3 provide the right ascension and declination (J2000 epoch), respectively. 
These coordinates were obtained from our astrometrized catalogs (see Sect. 2). Column 4 gives the type of RR Lyrae star, whether RRab or RRc, 
Columns 5 and 6 list the pulsation period and the Heliocentric Julian Day of maximum light, respectively. 
With the only exception of the period for star V2 (see discussion below), all periods are accurate to the fifth digit. 
Columns 7 and 8 give the intensity-weighted mean $B$ and $V$ magnitudes, while Columns 9 and 10 list the corresponding amplitudes of the light variation. 
Finally, Columns 11 and 12 provide metallicites for some of the variable stars measured either spectroscopically or photometrically (see Section 4).
 The light curves of the variable stars are presented in Figure~\ref{fig:curve}. The $B, V$ time-series data of each variable star
are provided in Table~3, which is published in its entirety in the electronic edition of the journal.
Three of the RR Lyrae stars identified in UMa~I (namely, V1, V2 and V3) lie inside the galaxy half light radius defined by \citet{mar08}, (see Section 5). 
Stars V4, V5 and V6 lie slightly outside this region, but are still inside the galaxy radius.
Finally, V7 is located at the edge of the FOV covered by our observations at a distance of $21.2\arcmin$ from the galaxy center.
Star V2 is close to a very bright red ($B-V =1.1$ mag) object, that significantly affects the star $V$ magnitudes making it rather 
difficult to interpret the visual light curve (see Figure~\ref{fig:curve}), whereas the $B$ light curve appears to be only marginally affected. 
For this reason the period of  V2 is less accurate than for the other 6 variables. 
Star V7 also has a scattered light curve.
Its periodicity and the position on the  HB (see Section 5) qualify V7 as 
a long period c-type RR Lyrae, but the star color, $B-V =0.10$ mag, appears to be much bluer than the first-overtone blue edge of the 
RR Lyrae instability strip, for which the canonical value is $B-V \sim 0.18$ mag (see, e.g, \citealt{walker98}). Further, the $V$-band amplitude of V7 is slightly larger than the  $B$ amplitude.
 The  blue color of V7, the scatter in the light curve and the reduced amplitude in the $B$ band may be explained with the star  being 
contaminated by a faint, blue, close companion.  Finally, V4 is about 0.15-0.2 mag brighter than the other RR Lyrae stars, this along with the 
rather long period  (0.75 days) may perhaps  indicate that the star is evolved off the Zero Age Horizontal Branch (ZAHB).

\subsection{BAILEY DIAGRAM AND OOSTERHOFF CLASSIFICATION}

\begin{figure*}
\centering
\includegraphics[scale=.80]{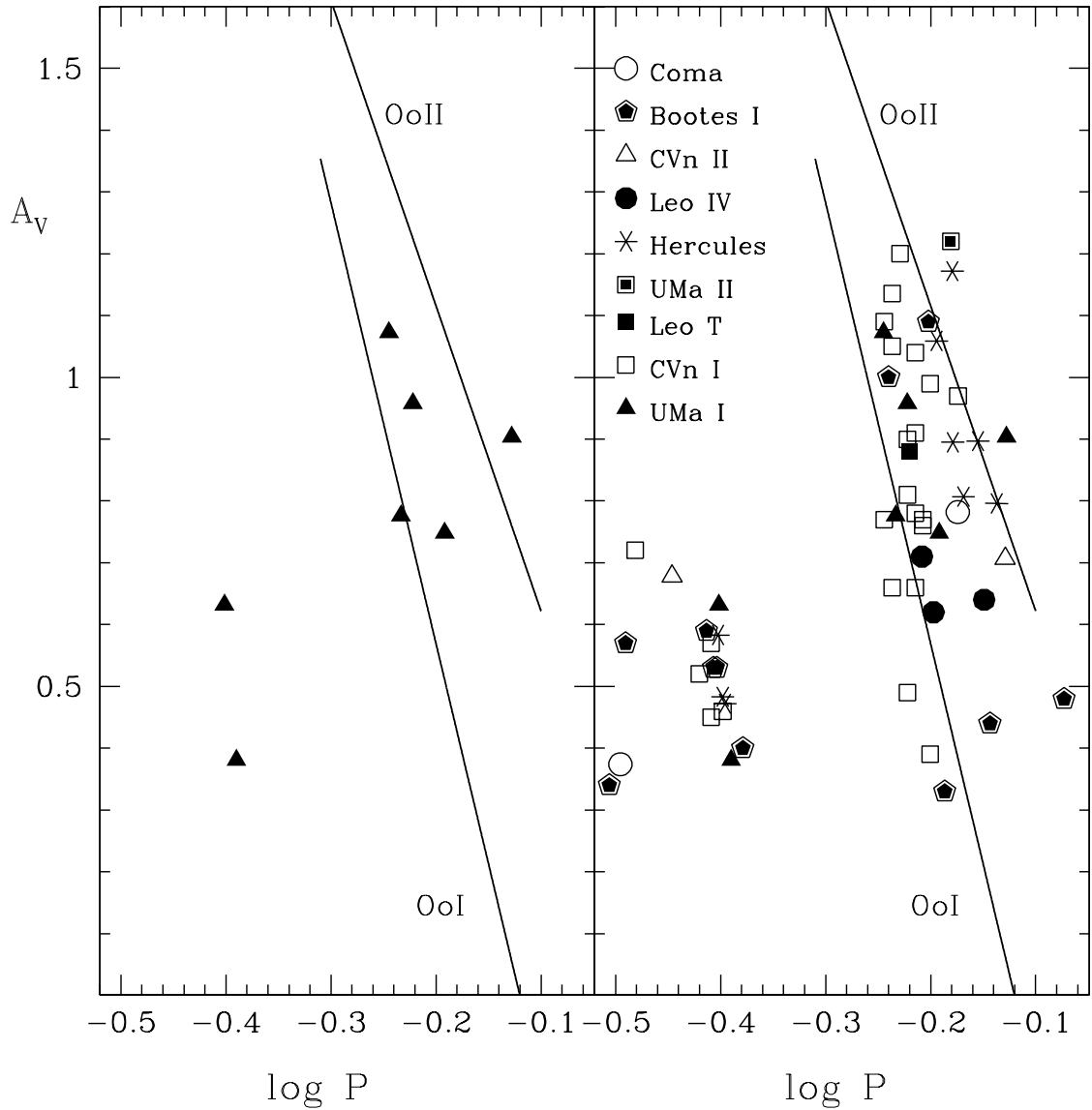}
\caption[]{\textit{Left}: $V$-band period-amplitude diagram of the RR Lyrae stars in UMa~I ; \textit{Right}:  Comparison with other eight UFDs studied for variability. In both panels the solid lines show the loci of the OoI and OoII Galactic GCs from \citet{cle00}.}
\label{fig:fig2}
\end{figure*}

Figure~\ref{fig:fig2} shows the $V$-band period-amplitude (Bailey) diagram of the UMa~I RR Lyrae stars
 (filled triangles), according to the periods and $V$ amplitudes reported in Table \ref{t:1}.
We have also plotted in the figure  the RR Lyrae stars identified in other 7 UFDs:
   Coma \citep{mus09}, Leo IV \citep{mor09}, Bootes I \citep{ora06}, CVn II \citep{gre08}, 
   Hercules \citep{mus12}, UMa~II \citep{ora12} and Leo T \citep{cle12}. 
The solid lines 
 show  the loci of the OoI and OoII Galactic GCs according to \citet{cle00}.
 The position of  the UMa~I  RR Lyrae stars in Figure~\ref{fig:fig2} does not permit a firm classification of  the galaxy in 
one of the two Oosterhoff groups. 
The average period of the five RRab stars in UMa~I  
is $\langle$P$_{\rm ab}\rangle$=0.628 days ($\sigma$=0.071 days, average on 5 stars). This value
and the metallicity from \citet{kir08,kir11}, locate UMa~I near the border 
between  Oo-intermediate (Oo-Int) and OII objects in the [Fe/H]-$\langle$P$_{\rm ab}\rangle$ 
plane (see, e.g., Figure 10 of \citealt{contreras13}). However, if we discard the longest period RRab variable (star V4, see Section 3) the average period becomes 
$\langle$P$_{\rm ab}\rangle$=0.599 days ($\sigma=0.032$ days,  average on 4 stars), and similarly if we only 
consider the RRab stars inside the galaxy half-light radius (V1, V2 and V3)  we obtain $\langle$P$_{\rm ab}\rangle$=0.599 days ($\sigma=0.039$ days, average on 3 stars) 
that definitely suggest an Oo-Int  classification for UMa~I.

\section{METALLICITY AND DISTANCE TO UMaI  FROM THE RR LYRAE STARS}

An estimate of the metal abundance of the  UMa~I RR Lyrae stars can be obtained exploiting 
 the relation existing between pulsation period and  $\phi_{31}$ parameter of the Fourier decomposition of the $V$-band light 
curves \citep[see][for a description of the method for RRab and RRc stars, respectively]{jur96,mrg07}. 
 Fourier parameters for the RR Lyrae stars we have identified in UMa~I are provided in Table~\ref{tab:Fourier_Parameters} along with the r.m.s. scatter (column 11) of the 
Fourier best fit of the light curves. The last column of the table also 
provides values of D$_m$, a parameter measuring the regularity of the light curve  for fundamental-mode RR Lyrae stars and setting the so-called compatibility condition, namely, D$_m <$3,  for a reliable application of \citet{jur96} method  to RRab stars.
Of the UMa~I  fundamental-mode RR Lyrae stars only V5 has D$_m <$3 (see Table~\ref{tab:Fourier_Parameters}). The metallicity of V5 derived with this technique
 is [Fe/H]=$-2.3 \pm0.2$ dex,  on the \citet{car09} metallicity scale (see column 12 of Table~\ref{t:1}). 
Among the other RRab stars V1,  V3  and V4 do 
not strictly satisfy the \citet{jur96} compatibility condition, but are at the limit (V1 and V3) or slightly above (V4)  the relaxed condition used in \citet{cacciari05} (D$_m <$5),
whereas  V2 has a  large D$_m$ value definitely exceeding both \citet{jur96} and \citet{cacciari05} conditions.  
Since the Fourier fit  of the light curves of V1, V3 and V4 appears to be very regular we still obtained an estimate of metallicity for these three stars, for which we find [Fe/H]$_{\rm V1}=-2.3 \pm 0.2$ dex,  [Fe/H]$_{\rm V3}=-2.2 \pm 0.3$ dex, and [Fe/H]$_{\rm V4}=-2.4 \pm 0.2$ dex. Our photometric metallicities  are in very good agreement with
the spectroscopic metallicity derived for V1, V3, and V4 by \citet{kir08}  (see  column 11 of Table~\ref{t:1}).
%
%
%
%
%
%
%
For the RRc stars we used the [Fe/H]-$\phi_{31}$-P relations derived by \citet{mrg07} obtaining [Fe/H]$_{\rm V6}$=$-2.3 \pm 0.1$ dex, and [Fe/H]$_{\rm V7}$=$-2.2 \pm 0.2$ dex,  
on the \citet{car09} metallicity scale. 
The mean metallicity of the UMa~I RR Lyrae stars derived from the Fourier parameters of the light curves is then [Fe/H]=$-$2.29 dex ($\sigma$=0.06 dex, average on 6  stars), 
 in excellent agreement  with the mean metallicity of the UMa~I red giants measured by \citet{kir08} and only 0.1 dex lower than the revised value by \citep{kir11}.

The mean magnitude of the RR Lyrae stars, $\langle V(RR) \rangle$, can be used to derive the distance to UMa~I. We find  $\langle V(RR) \rangle$=$20.40\pm0.08$ mag  as the average over the 7 stars.
However, as discussed in Section 3, star V4 appears to be about 0.15-0.20 mag
brighter than the other RR Lyrae stars and if, as we suspect, the star is evolved off the ZAHB it should not be used to estimate the galaxy distance. Furthermore, as UMa~I appears to be rather elongated and
structured (see Section 5), it may not be appropriate to average  all together the RR Lyrae stars we have identified in the FOV of our observations   to estimate the distance to the center of the galaxy.  
To have a better insight into this issue we computed the mean  $\langle V(RR) \rangle$ value using different selections of the RR Lyrae and have summarized our results in Table~4. 
The different $\langle V(RR) \rangle$ estimates (see column 2 of Table~4) agree within the errors and are all consistent with the $V$ magnitude of the galaxy HB ($V(HB) = 20.45 \pm 0.02$ mag) derived 
by \citet{oka08}, thus confirming the reliability of our absolute photometric calibration.
However, the dispersion of $\langle V(RR) \rangle$ is significantly reduced  if star V4 is discarded, therefore in the following we adopt the mean value obtained 
discarding star V4,  $\langle V(RR) \rangle$= 20.43$\pm$ 0.02 mag,  to estimate the distance 
to UMa~I from the galaxy's RR Lyrae stars.
We have also assumed M$_{\rm V}=0.54\pm0.09$ mag  for the absolute visual magnitude of RR Lyrae stars with metallicity of [Fe/H] = $-$1.5 dex \citep{cle03} and 
$\frac{\Delta {\rm M_V}}{\Delta {\rm[Fe/H]}}=0.214\pm0.047$ mag/dex \citep{cle03,gratton04} for the 
slope of the luminosity-metallicity relation of the RR Lyrae stars. For the galaxy's metallicity we have adopted the mean value we have derived from the RR Lyrae stars  [Fe/H]=$-2.29\pm0.03$ dex.
\citet{sc98} reddening maps give $E(B-V)$=0.019 $\pm$ 0.026 mag in the direction of UMa~I, however, we obtain a slightly higher value of $E(B-V)$=0.04 $\pm$ 0.02 mag
by fitting the galaxy CMD to the ridge-lines of the Galactic GC M68 (see Section 5). This higher value is in excellent agreement with the reddening that \citet{bro12} 
obtain by fitting UMa~I's  RGB to the ridgeline of the GC M92. In the following we have thus assumed $E(B-V)$=0.04 $\pm$ 0.02 mag and the standard extinction law: A$_V$=3.1$\times E(B-V)$.
The distance modulus of UMa~I  derived with this procedure is $\mu_0$ = $19.94 \pm 0.13$ mag, corresponding to  a distance d=$97.3^{+6.0}_{-5.7}$ kpc, where 
the errors  include the contribution of the uncertainties in the metallicity, reddening, photometry, photometric calibration, slope and zero point of the M$_{\rm V}$ {\it  vs} [Fe/H] relation, and on the average apparent visual magnitude of the UMa~I  RR Lyrae stars.
Our distance modulus is in very good agreement with the distance modulus of  $\mu_0$ = $19.93\pm0.10$ mag, obtained by \citet{oka08} using  
the $V$ magnitude of the HB estimated from the comparison with M92. Our value is also  consistent, within the errors, with the distance modulus $\mu_0$ = $19.99\pm0.04$ mag obtained 
for UMa~I by \citet{bro12}. 
 Finally, we note that if we assume for the absolute visual magnitude of the RR Lyrae stars  the brighter zero point by \citet{Ben11}, M$_{\rm V} = 0.45 \pm  0.05$ mag, for  [Fe/H] = $-$1.5 dex, we obtain 
 $\mu_0$=  20.03 mag,   which is respectively 0.10 mag and 0.04 mag longer than derived by \citet{oka08} and \citet{bro12}.
 
\section{CMD AND SPATIAL DISTRIBUTION OF UMa~I  MEMBER STARS}

\begin{figure*}[t!]
\begin{tabular}{c}
\multicolumn{1}{c}{\includegraphics[width=16.5cm,height=13.5cm]{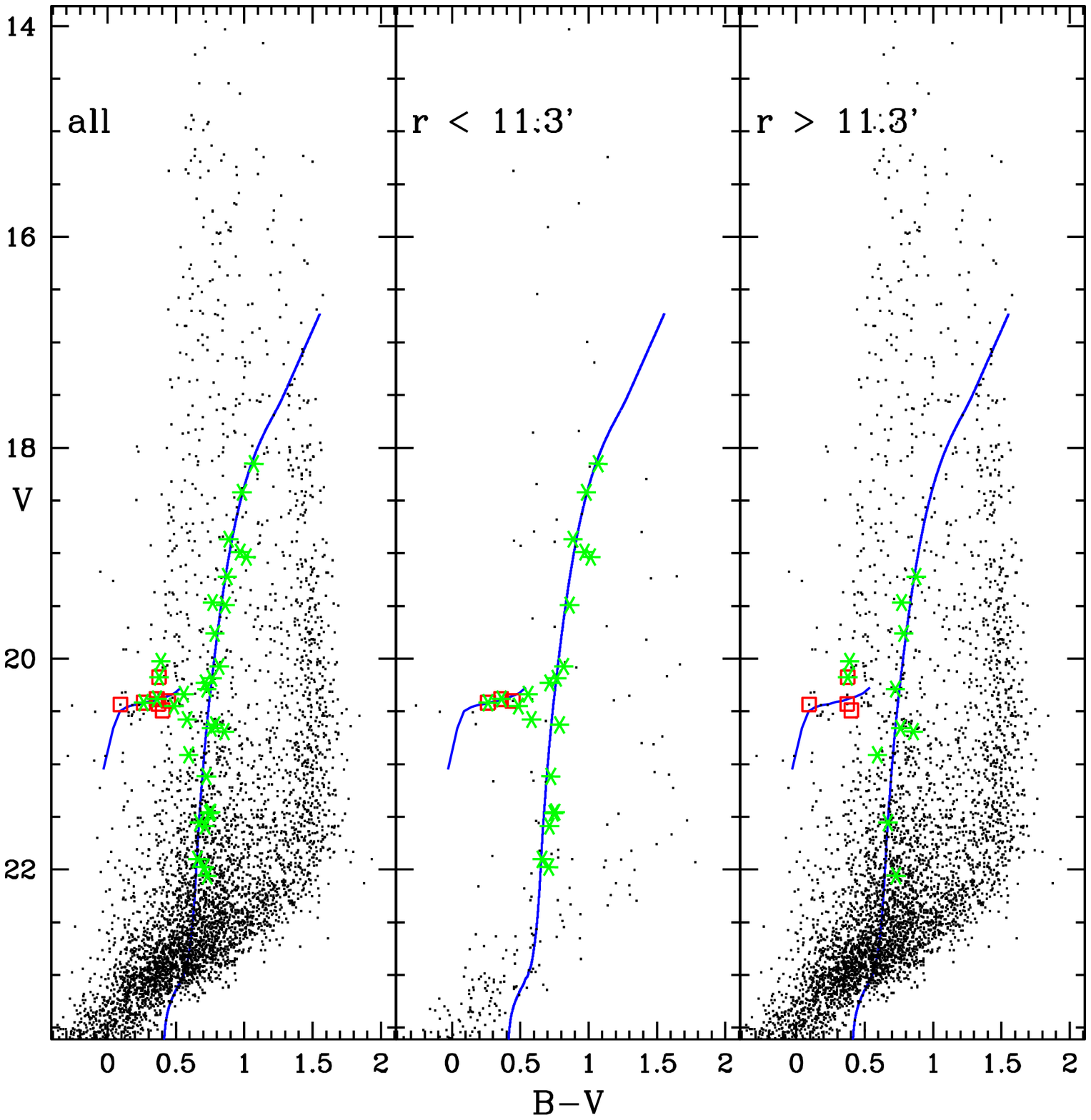}}\\
\multicolumn{1}{c}{\includegraphics[width=16.5cm,height=4.5cm]{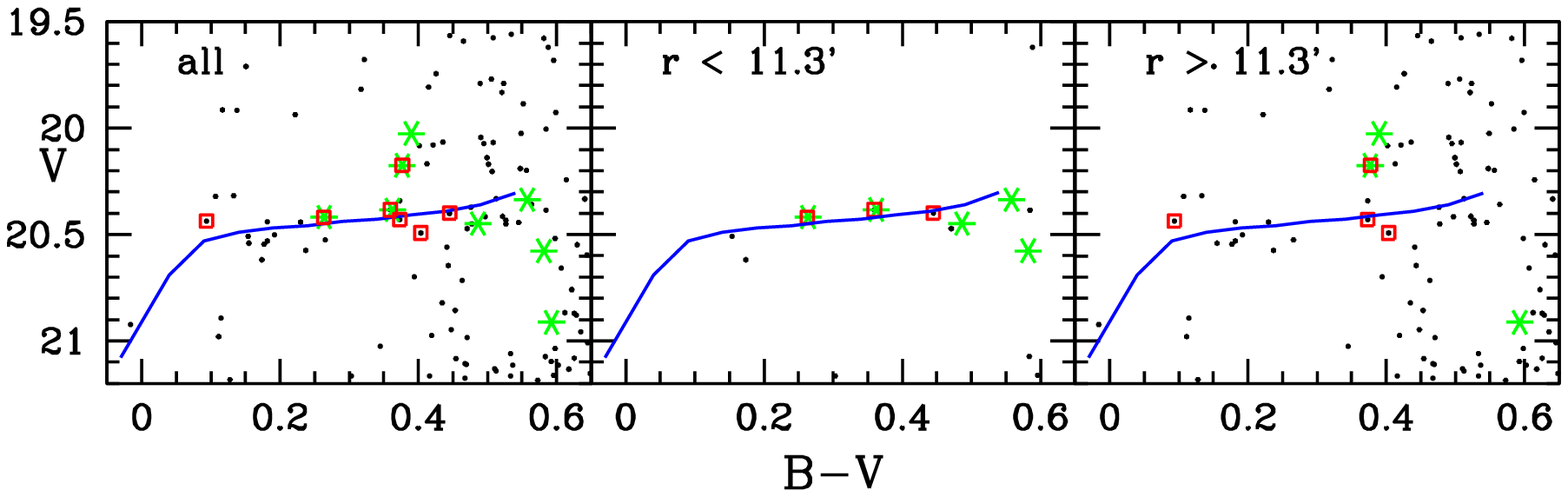}}\\
\end{tabular}
\caption{\textit{Upper panel - left}: $V$, $B-V$ CMD of all the stellar detections in the FOV of the TLS telescope   
 with valid photometry on
 all input images, global Sharp parameter $-0.4 \leq$ Sharp $\leq0.4$, and $\chi < 1.7$, in each filter. 
 The RR Lyrae stars are marked with (red) squares. Member stars of UMa~I 
 confirmed spectroscopically by \citet{sim07} and Kirby et al. (2008) are marked with  (green) 
 asterisks. The (blue) solid line is the ridge line of the Galactic GC M68.
\textit{Upper panel -  middle}: Same as in the left panel, but for stars in the half-light radius of UMa~I. 
  \textit{Upper panel - right}:
 Same as in the left panel, but for stars outside the galaxy half-light radius.  \textit{Lower panels}: Zoomed-in view centered on the  HB region of  the CMDs shown in the upper panels. 
 }
\label{fig:fig3}
\end{figure*}

The $V$, $B-V$ CMD of UMa~I obtained in this study is shown in Figure~\ref{fig:fig3}. We used the quality information provided by
 the  \texttt{DAOPHOT-ALLSTAR-ALLFRAME} packages, namely, the  \texttt{ALLFRAME} parameters $\chi$ and Sharp
\citep[see ][]{ste92}, to clean the list of detected sources, selecting for the CMD only stellar detections with valid photometry on
 all input images, global Sharp parameter $-0.4 \leq$ Sharp $\leq0.4$, and $\chi < 1.7$, in each filter. 
 This allowed us to reduce the contamination by background galaxies.
In the left panel of   Figure~\ref{fig:fig3} we have plotted all the stellar detections in the FOV of the TLS telescope, whereas in the middle panel 
 we have reported only stars within the
 galaxy's half-light radius \citep[r$_h$ = 11.3 $\arcmin$,][]{mar08}. Since UMa~I is rather elongated, the radius used for the selection 
was defined by the equation 
 $r^2 = {{x}^{2} + {y}^{2}/{(1-e)^{2}}}$ 
where $e$ is UMa~I's ellipticity for which we used $e$=0.80 from \citet{mar08}. 
Finally, the right panel shows instead objects in the TLS FOV that are located outside UMa~I's  half-light radius. 
In all panels the RR Lyrae stars are marked by (red) open squares. 
 Typical internal errors of the combined photometry for non-variable stars at the magnitude level of UMa~I HB
(V $\sim20.5$ mag) are $\sigma_{V}=0.017$ mag and $\sigma_{B}=0.026$ mag, respectively.
The faintest stars in our CMD have V $\sim23$ mag (at a signal-to-noise ratio of $\sim$ 6).  
In order to distinguish stars belonging to the UMa~I galaxy  from the overwhelming population of field stars belonging to the MW halo and disk we 
 have cross-matched our photometric catalog against  \citet{sim07} and \citet{kir08} catalogs of spectroscopically confirmed UMa~I members. 
There are 32 stars in common between these  catalogs, they are marked  as (green) asterisks in Figure ~\ref{fig:fig3}. Three of them  are RR Lyrae 
stars identified in the present study, namely, stars V1, V3 and V4, whereas the remaining stars are mainly red 
giants. These stars  along with the RR Lyrae variables allow to identify quite clearly UMa~I's  red giant and horizontal branches. We have then  
 overlaid on
 the CMDs in Figure~\ref{fig:fig3} as (blue) solid lines the mean ridge-lines of 
the Galactic GC  M68 taken from \citet{wal94}. This cluster was selected because its 
 metal abundance, ($[Fe/H]_{M68} = -2.27 \pm 0.04$ dex, \citealt{car09}), well matches the metallicity of UMa~I.
The cluster ridge lines were shifted by +4.79 mag in $V$ and $-$0.01 mag in $B-V$  to match the galaxy HB, the RR Lyrae stars, and the red giants 
with spectroscopically confirmed membership. 
These shifts imply a reddening $E(B-V)$ = 0.04$\pm$0.02 mag for UMa~I, in excellent agreement with \citet{bro12}.
 The comparison with M68 confirms that UMa~I, like M68 has a mostly old and metal-poor stellar population. The mean magnitude of the galaxy HB 
inferred from the match to M68 is $V = 20.45 \pm 0.11$ mag  in excellent agreement with \citep{oka08}, and very well consistent with  the average magnitude of the RR Lyrae stars 
identified in this work, especially if the brightest and likely evolved of our variables (star V4) is discarded. 

\begin{figure*}
\centering
\includegraphics[scale=.85]{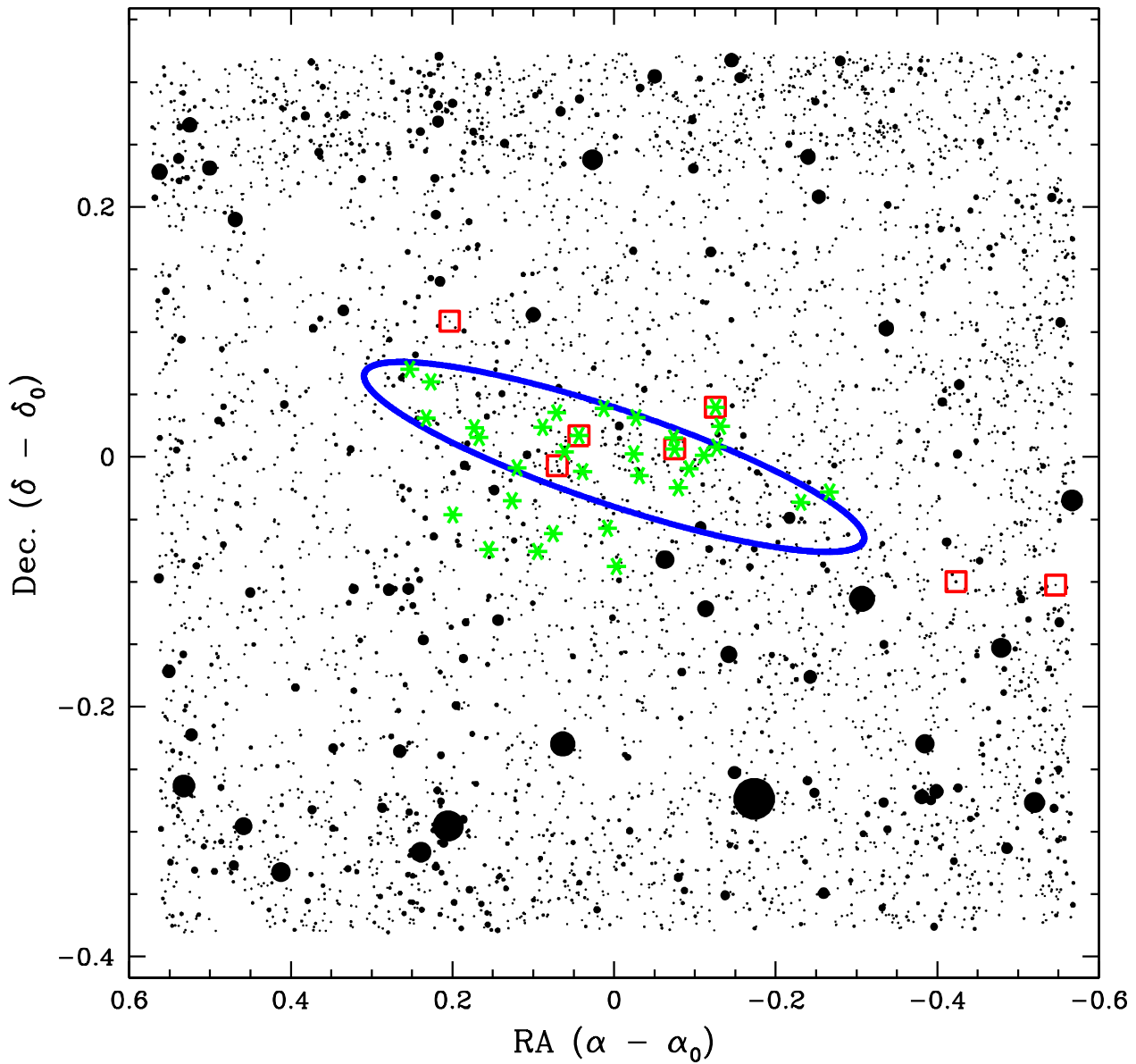}
\caption{Map of all the sources 
in the FOV of the TLS telescope that we plot  in the CMD in Figure~\ref{fig:fig3}. Symbol sizes are inversely proportional to the objects' magnitude. The (blue) ellipse traces the half-light radius region of UMa~I, according to the position angle, half-light radius, ellipticity and center coordinates ($\alpha_0$, $\delta_0$) of \citet{mar08}.
The (red) squares are the RR Lyrae identified in the present study. The (green)  asterisks are UMa~I members spectroscopically confirmed by \citet{sim07} and \citet{kir08}. 
}
\label{fig:fig4}
\end{figure*}

\begin{figure*}
\centering
\includegraphics[scale=.80]{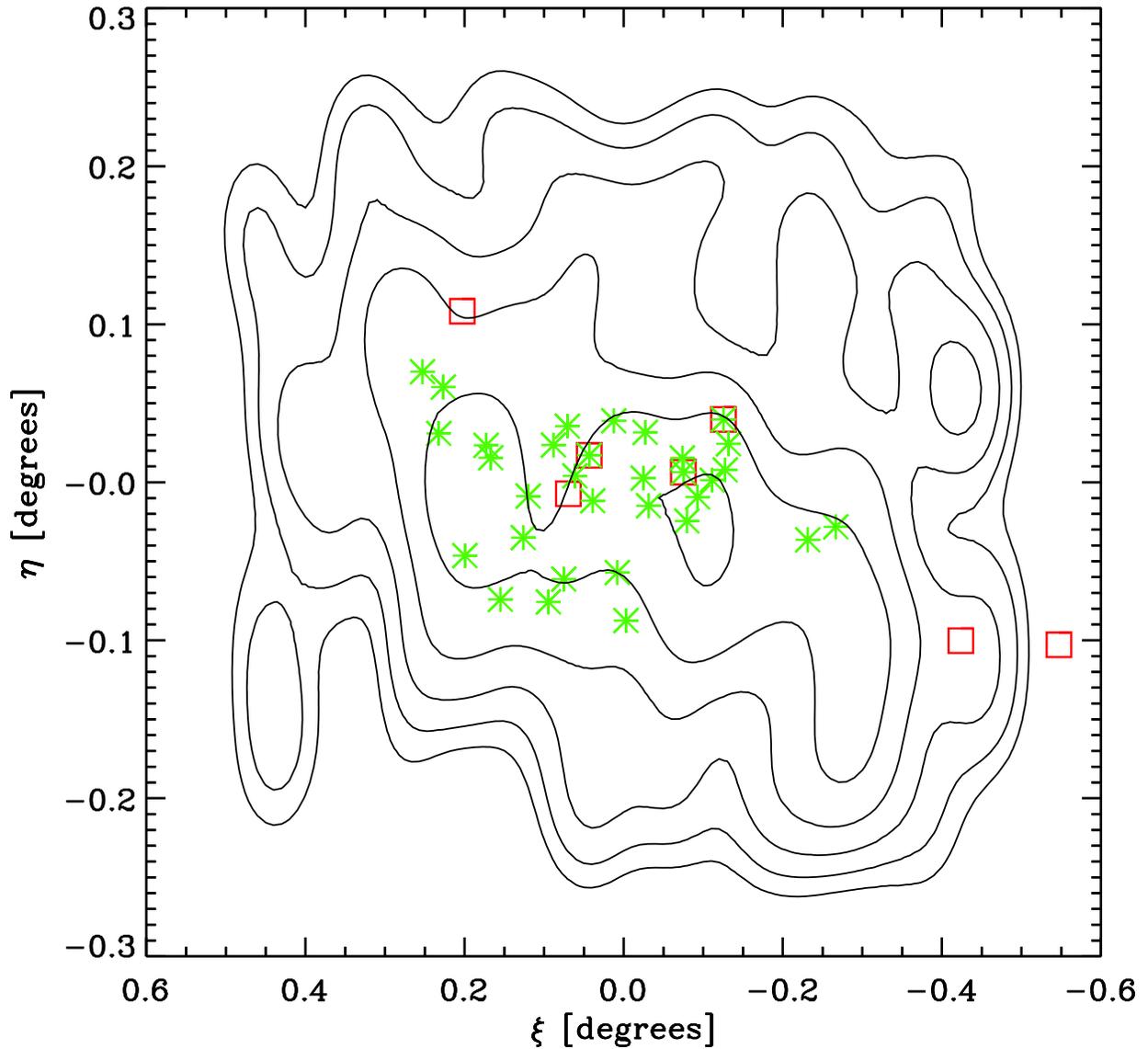}
\caption{Isodensity contours of UMa~I RGB and HB stars selected from the galaxy CMD by considering only stars within 0.1 mag from the ridge lines
of M68 (see text for details). 
The plotted contour levels are from 3 $\sigma$ to 15 $\sigma$ above the background level.
The (green)  asterisks are UMa~I members spectroscopically confirmed by \citet{sim07} and \citet{kir08}, the 
(red) open squares are the RR Lyrae stars identified in this study.
}
\label{fig:fig5}
\end{figure*}

Figure ~\ref{fig:fig4} shows a map of the stellar sources in the FOV of the TLS telescope that are in the CMD in the left panel of Figure~\ref{fig:fig3}. The (blue) ellipse is drawn
 using the half-light radius, the ellipticity and the position angle of UMa~I derived by \citet{mar08}.  Same as  in Figure~\ref{fig:fig3} the (green) 
asterisks are UMa~I member stars spectroscopically confirmed by \citet{kir08} and \citet{sim07}, while the open (red) squares are the RR Lyrae identified in this work. The latter appear 
to be distributed  along the galaxy major axis.
In Figure~\ref{fig:fig5} we show the isodensity map of RGB and HB stars selected from the CMD of UMa~I (see Figure~\ref{fig:fig3})
by considering only stars within 0.1 mag from the ridge lines
of M68. These stars were binned in $2.4'\times2.4'$ boxes
and smoothed by a Gaussian kernel of full-with at half maximum of $2.4'$.
The contours level are from 3 $\sigma$ above the background. 
The spectroscopically confirmed members of UMa~I, the RR Lyrae stars, as well as 
the central isodensity contours appear to trace an S-shaped structure. This shape is characteristic 
of  dSph galaxies undergoing tidal stripping 
\citep[e.g. HCC-087, ][]{koc12}. In conclusion, both the high value of the 
ellipticity and the characteristic S-shaped structure of the stellar distribution suggest that UMa~I is  tidally interacting with the MW.

\section{SUMMARY AND CONCLUSIONS}

We have performed the first study of the variable stars in the UMa~I UFD and 
detected seven RR Lyrae stars in the galaxy, of which five are RRab 
and two are RRc pulsators. Three of the RR Lyrae stars we have identified in UMa~I were independently 
classified as member stars by \citet{kir08} using medium-resolution spectroscopy.
The average period of the five RRab stars is $\langle P_{\rm ab} \rangle$ =$0.628$ days (or  $\langle P_{\rm ab} \rangle$ =$0.599$, if the brightest and longest period variable, star V4, is discarded)  and
suggests an Oo-Int classification for this UFD. 
UMa~I is not the first of the new MW satellites to be classified as Oo-Int. The RR Lyrae stars identified in Canes Venatici~I  (CVn~I; \citealt{kue08}) clearly have Oo-Int properties, and similarly, the
lone RR Lyrae star identified in Leo~T \citep{cle12} also suggests an Oo-Int classification for that UFD. However, both  CVn~I and Leo~T have characteristics that set them apart from the 
 ``bona-fide" MW UFDs. Specifically,  CVn~I  is the brightest of the new MW satellites, and the high luminosity makes this galaxy much more similar to the classical  dSphs than to the UFDs (see Figure~1 in \citealt{cle12}). On the other hand, Leo~T is the only UFD found to contain a significant amount of neutral gas  and the lowest luminosity galaxy with ongoing star formation known to date. 
Hence, UMa~I  is so far the only ``bona-fide" UFD to exibit  Oo-Int properties.
The distance modulus of UMa~I derived from the mean $V$ magnitude of the RR Lyrae stars is $\mu_0$ = $19.94\pm0.13$ mag, corresponding to  d=$97.3^{+6.0}_{-5.7}$ kpc. 
This distance is in good agreement with the estimates by \citet{oka08} and  \citet{bro12}. 
The individual metallicities derived for the RR Lyrae stars  from the $\phi_{31}$ parameter of the Fourier decomposition of the light curve are in good agreement 
with the average spectroscopic metallicity derived for UMa~I red giants by \citet{kir08,kir11}.
The isodensity contours  of red giants and HB stars properly selected from the  galaxy CMD, and the spatial distribution of both RR Lyrae stars and spectroscopically confirmed 
red giants (\citealt{sim07, kir08}) show that UMa~I  has an S-shaped structure, typical of 
dwarf galaxies undergoing tidal interaction.



\acknowledgments
We thank an anonymous referee, for comments and suggestions that helped to improve the manuscript.
We warmly thank P. Montegriffo for the development and maintenance of the GRATIS software,  and G. Battaglia for providing the routines of the isodensity map.
Financial support for this research was provided by COFIS ASI-INAF I/016/07/0 and by PRIN INAF 2010 (P.I.: G. Clementini).




\begin{table*}
\caption[]{Parameters of the Fourier decomposition of  the $V$-band light curve}
\begin{tabular}{llcccccccccc}
\hline
\hline
Star  & Type &$ \phi_{21}$& $err\phi_{21}$ & $\phi_{31}$& $err\phi_{31}$& $A_{21}$& $errA_{21}$  &  $A_{31}$&  $errA_{31}$  &  $\sigma$& D$_m$         \\
\hline
V1    & RRab &  2.44    & 0.08	 &   4.4~    & 0.1~         & 0.39  & 0.03       & 0.28   &  0.03       &  0.05 & 5         \\  
V2    & RRab & 2.6~     & 0.2~	 &   5.3~    & 0.6~         & 0.44  & 0.09       & 0.18   &  0.07       &  0.1~ & 11        \\  
V3    & RRab & 2.3~     & 0.2~	 &   4.8~    & 0.2~         & 0.23  & 0.04       & 0.25   &  0.04       &  0.07 & 5          \\  
V4    & RRab & 2.28     & 0.08	 &   5.06    & 0.09         & 0.52  & 0.04       & 0.45   &  0.04       &  0.05 & 6         \\  
V5    & RRab & 2.16     & 0.06	 &   4.5~    & 0.07         & 0.48  & 0.03       & 0.36   &  0.02       &  0.03 & 2.7      \\  
V6    & RRc    & 2.8~     & 0.2~	 &   5.6~    & 0.5~         & 0.29  & 0.05       & 0.12   &  0.05       &  0.06 & (a)                  \\  
V7    & RRc    & 3.4~     & 0.4~	 &   6.6~    & 0.9~         & 0.4~  & 0.1~	      & 0.2~   &  0.1~        &  0.07 & (a)                    \\  
\hline
\end{tabular}
\label{tab:Fourier_Parameters}
\begin{flushleft}
(a) \footnotesize{The D$_m$ value is defined only for RRab stars}
\end{flushleft}
\end{table*}

\begin{table*}
\caption[]{Identification and properties of the RR Lyrae stars identified in UMa~I}
\footnotesize
\label{t:1}
\begin{tabular}{l c c l l c c c c c c c c c}
\hline
\hline
\noalign{\smallskip}
 Name & $\alpha$ &$\delta$ & Type & ~~~P  & Epoch (max)        & $\langle B \rangle$ & $\langle V \rangle$ & A$_{B}$ & A$_{V}$ &[Fe/H] &[Fe/H]\\
 	    &	(2000)   & (2000)   &	         &~(days)& JD ($-$2454000) & (mag)                       & (mag)                         & (mag)     &  (mag)     &  (a)    &  (b)                            \\
 	    \noalign{\smallskip}
	    \hline
	    \noalign{\smallskip}
	    
V1  & 10:34:59.2  & +51:57:07.3   & RRab  &  0.56924      &  915.455  & 20.72   &  20.47   & 1.29 & 1.07 &$-$2.16 & $-2.3 \pm 0.2$ \\   
V2  & 10:35:05.5  & +51:55:39.8   & RRab & 0.584             &  886.660  & 20.85   & 20.37   & 1.09 & 0.78  &      -       &   -          \\	  
V3  & 10:34:30.8  &   +51:56:28.9 & RRab & 0.64315        &  886.225  & 20.73   &  20.40  & 0.99 & 0.75  &$-$2.17 &  $-2.2 \pm 0.3$ \\
V4  & 10:34:18.7  &   +51:58:29.2 & RRab & 0.74516        &  942.400  &  20.62  &  20.24  & 1.15 & 0.90  &$-$2.30 &  $-2.4 \pm 0.2$\\
V5  & 10:35:37.5  &  +52:02:35.6  & RRab &  0.59967       &  916.498  & 20.87   &  20.49  & 1.38 & 0.96  &     -         &  $-2.3 \pm 0.2$\\ 
V6  & 10:33:07.1  & +51:50:05.1   & RRc  & 0.39673         &   886.980  & 20.73  &  20.42  & 0.72 & 0.63   &-              &$-2.3 \pm 0.1$ \\ 	 
V7  & 10:32:37.5  & +51:49:55.7   & RRc  & 0.40749         &   947.300  &  20.54 &  20.44  & 0.37 & 0.38   &-              &$-2.2 \pm 0.2$ \\  

\hline
\end{tabular}
Notes:\\
 (a) Metallicities from \citet{kir08}\\
 (b) Photometric metallicities from the Fourier parameters of the $V$-band light curve, they are on the \citet{car09} metallicity scale. Note that of the RRab stars only  V5 fully satisfies the  \citet{jur96} compatibility condition (see text for details). 
\normalsize
\end{table*}

\begin{table*}
\begin{center}
\caption[]{Johnson-Cousins $B$, $V$ photometry of UMa~I variable stars.}
\label{t:2}
\begin{tabular}{l c c c c c c}
\hline
\hline
\noalign{\smallskip}
\multicolumn{4}{c}{Star V1 - RRab}\\
\noalign{\smallskip}
\hline
\noalign{\smallskip}
HJD & $B$ &$err_B$& HJD & $V$ &$err_V$\\
(-2454940) & (mag) & (mag) & (-2454940) & (mag) & (mag)\\
\noalign{\smallskip}
\hline
\noalign{\smallskip}
1.33650 & 21.11 & 0.05 & 1.36133 & 20.79 & 0.04 \\
1.39005 & 21.26 & 0.05 & 1.41162 & 20.96 & 0.06 \\
1.43730 & 21.24 & 0.05 & 1.45850 & 20.85 & 0.06 \\
1.46563 & 21.16 & 0.05 & 1.48730 & 20.76 & 0.05 \\
2.35583 & 20.82 & 0.03 & 2.38037 & 20.50 & 0.04 \\
2.39971 & 21.05 & 0.04 & 2.42432 & 20.62 & 0.05 \\
2.43562 & 21.14 & 0.05 & 2.46143 & 20.64 & 0.04 \\
3.33411 & 19.98 & 0.04 & 3.35889 & 19.88 & 0.05 \\
3.40212 & 20.34 & 0.04 & 3.42725 & 20.19 & 0.06 \\
3.43734 & 20.54 & 0.03 & 3.46191 & 20.33 & 0.06 \\
\hline
\end{tabular}
\end{center}
This table is published in its entirety in the electronic edition of the journal. A portion is show here for guidance regarding its form and content.
\normalsize
\end{table*}

\begin{table*}
\begin{center}
\caption[]{Average magnitude of UMa~I RR Lyrae stars according to different sample selections.}
\label{t:3}
\begin{tabular}{l c c  l}
\hline
\hline
\noalign{\smallskip}
Star's selection & $\langle V(RR) \rangle$ & $\sigma$&Notes\\
                            & (mag)                                 &  (mag)      &\\
\noalign{\smallskip}
\hline
\noalign{\smallskip}
All                       &  20.40$\pm$0.03& 0.08 &   \\
All minus V4     &  20.43$\pm$0.02& 0.04 &   \\
V1+V2+V3        &  20.41$\pm$0.03& 0.05 & Only stars inside the r$_{\rm h}$ \\
V4+V5+V6+V7 &  20.40$\pm$0.06& 0.11 & Only stars outside the r$_{\rm h}$ \\
V5+V6+V7        &  20.45$\pm$0.04& 0.03 & Only stars outside the r$_{\rm h}$ and excluding V4 \\ 
\hline
\end{tabular}
\end{center}
\normalsize
\end{table*}

\end{document}